\documentclass[aps,prl,amsmath,amssymb,preprint]{revtex4}
\usepackage{graphicx}
\usepackage{amsmath}
\usepackage{amssymb}
\usepackage{color}
\usepackage{soul}

\begin{document}

\title{Probing equilibrium glass flow up to exapoise viscosities}

\author{E.A.A. Pogna$^{1}$,
C. Rodr\'iguez-Tinoco$^{2}$,
G. Cerullo$^{3}$,
C. Ferrante$^{1}$,
J. Rodr\'iguez-Viejo$^{2,4}$,
T. Scopigno$^{1}$
}

\affiliation{$^{1}$Dipartimento di
Fisica,~Universit\'a~di~Roma~"La Sapienza",~I-00185,~Roma,~Italy \\
$^{2}$Nanomaterials and Microsystems Group, Physics Department,
Universitat Aut$\grave{o}$noma de Barcelona,~ES-08193,~Bellaterra,~Spain\\
$^{3}$IFN-CNR, Dipartimento di Fisica,~Politecnico di Milano,~I-20133,~Milano,~Italy\\
$^{4}$MATGAS 2000 AIE, Campus UAB, E-08193, Bellaterra, Spain
}





\begin{abstract}
{Glasses are out-of-equilibrium systems aging under the crystallization threat. During ordinary glass formation, the atomic diffusion slows down rendering its experimental investigation impractically long, to the extent that a timescale divergence is taken for granted by many. We circumvent here these limitations, taking advantage of a wide family of glasses rapidly obtained  by physical vapor deposition directly into the solid state, endowed with different ''ages'' rivaling those reached by standard cooling and waiting for millennia. Isothermally probing the mechanical response of each of these glasses, we infer a correspondence with viscosity along the equilibrium line, up to exapoise values. We find a dependence of the elastic modulus on the glass age, which, traced back to temperature steepness index of the viscosity, tears down one of the cornerstones of several glass transition theories: the dynamical divergence. Critically, our results suggest that the conventional wisdom picture of a glass ceasing to flow at finite temperature could be wrong.}
\end{abstract}
\maketitle



\section{significance}
Does the glass cease to flow at some finite temperature?" Answering this question -of pivotal importance for glass formation theories- would require ridiculously long observation times. We circumvent this infeasibility relating the (directly inaccessible) ultra-viscous flow of a liquid to the elastic properties of the corresponding glass, which we measure as function of its age. The older is the glass the lower is the temperature at which viscosity can be determined. Taking advantage of physical vapor deposition, we rapidly obtain a wide spectrum of ages rivaling those of millenary ambers, enabling viscosity determinations at values as large as those pertaining to the asthenosphere. Our result ultimately rules out the finite temperature divergence of the molecular diffusion timescale in a glass.

\section {introduction}
A wide range of liquids can be cooled below the freezing point avoiding crystallization. Under such circumstances, viscosity increases dramatically, up to the point where the system falls out-of-equilibrium \cite{angell_rel,angell_sup,BANGALORE,kai}. Whether \cite{woly,woly2,douglas_div,richert_div,wagner_div,richert_div1} or not \cite{jeppe_natphys,dimarzio_nondiv,ngai_nondiv,ngai_nondiv2,chang_nondiv,chang_nondiv1} this increase signals a divergent behaviour is a question hitting at the core of any glass transition theory, and has attracted an increasing number of experimental investigations \cite{mcK,oconnel,zhao_mckenna,mckenna_jcp,mckenna_comment,simon,simon2,simon3,simon4,simon5,brian}.

The thermal history of a liquid engraves the properties of the resulting, amorphous material (glass), which continuously hikes down a rugged energy landscape to enhance its stability in search of lower minima \cite{flying,sastry}.
From an opposite perspective, the way the equilibrium properties of a unique liquid are encoded in the peculiarities of different corresponding glasses, each endowed with different histories and stabilities, has always puzzled scientists and technologists \cite{BANGALORE,newyork}.
The fictive temperature $T_f$, for instance, unambiguously signals the out-of-equilibrium arrested state accompanying the glass transition \cite{tul}, and increases with the cooling speed, which is upper limited by the finite thermal conductivity of the sample. Upon slowing down the quenching rate, the system gets trapped in lower potential energy landscape basins (inherent structures), but crystallization is also facilitated. In an attempt to rationalize such phenomenology, one way to define the glass transition temperature is as the fictive temperature at a conventional cooling rate: $T_g=T_f(-10$ K/min$)$. Correspondingly, the system falls out-of-equilibrium over a timescale of $\approx$ 100 s, attaining viscosities of $10^{11} \div 10^{13}$ poise. The impact of such an arbitrariness is mitigated by the observation that, for bulk materials, ordinary cooling rates are often confined in the range -0.1 to -10 K/min and $T_f$ varies very little. Remarkably, this severely limits the free energy region accessible by the glass, at least on timescales useful for many potential applications.
It has been recently shown \cite{kearns_hiking,ediger_science,leon,flying}, however, that the limitations in exploring the bottom of the energy landscape can be circumvented synthesizing glasses directly well below $T_g$, avoiding the supercooling route by physical vapor deposition (PVD). These glasses are referred to as ultrastable, for their enhanced resistance against crystallization, and populate free energy regions as low as those of semi-fossilized resins \cite{mcK} aged for thousands of years.

Indomethacin (IMC), selected for this study, is a prototype organic glass-former endowed with antipyretic properties, which, in its crystalline form, has poor water-solubility and bioavailability \cite{hancock_characteristics_1997}, limitations partially removed in the amorphous phase. Obtaining ultrastable amorphous IMC on laboratory timescales does not improve water solubility, but enables a unique opportunity to approach the so called "ideal glass", i.e. an amorphous phase with the same entropy of the undelying crystalline form.
\section{Results}
Glass stability can be quantified by determining $T_f$ from the endothermic jump exhibited by the specific heat upon fast heating \cite{moy}, due to the transformation of the glass into supercooled liquid (Materials and Methods).
We present in Fig. 1 an assessment of the stabilities of IMC glasses prepared by PVD at six different substrate temperatures $T_{sub}$ in the range 190 K $\div$ 290 K, as determined using differential scanning calorimetry (DSC).
The $T_{f}$'s of the depositions testifies to their stability enhancement (compared to conventional glasses), which increases with the substrate temperature up to a maximum gap of 34 K with respect to $T_g$, achieved for the most ultrastable, lowest $T_{f}$ case ($T_{sub}$=266 K, in agreement with \cite{ediger_science}). The reason for such stability enhancement, attained by tuning $T_{sub}$, has been traced back to the increase of the surface mobility, which is believed to allow molecules to arrange themselves into lower energy configurations \cite{kearns_hiking,ediger_science,flying}. Specifically, for a given deposition rate, coming from the high temperature side, the (relatively large) surface mobility grants that the glass is "assembled" in equilibrium with its liquid, i.e. each layer of molecules is equilibrated before new molecules arrive in the deposition process. Within this regime, the lower is the temperature, the lower the entropy of the system and the higher the stability. Below a given temperature the mobility is so slow that the structure is no longer in equilibrium with the liquid, and the system becomes increasingly unstable

Since the shape of the inherent structures of the energy landscape defines the vibrational properties, a question arises on the possible connection between mechanical response and glass stability, in connection with recent experimental evidences in amorphous depositions and naturally aged glasses \cite{kearns_high-modulus_2010,mcK,ramos_velocity}. Specifically, previous studies \cite{fakharai} of IMC obtained by PVD reveal a sound velocity dependence on local substrate temperature which, in turn, has been shown to relate to the onset temperature of enthalpy release measured in calorimetry upscans \cite{fak_elips}. To address this issue we first studied the acoustic properties of the most ultrastable deposition by a combination of time domain pump-probe optical spectroscopy and frequency domain Inelastic X-ray Scattering (IXS). The former technique is a broadband version \cite{apl,natcom} of picosecond photo-acoustics (BPA) \cite{thomsen}, which we recently developed to efficiently generate and detect longitudinal acoustic wave-packets in sub-$\mu$m thick transparent layers.
Specifically, by using white light pulses ($450\div650$ nm), the energy dispersion and mode attenuation of the vibrational excitations have been simultaneously investigated in the frequency range $10\div20$ GHz. IXS, on the other hand, allows the study of the high frequency limit (THz), where the characteristic excitation wavelengths approach the average interparticle distance and the structural disorder becomes of crucial importance.
Further details on both the techniques are given in Materials and Methods.
The dispersion curve as determined by BPA in the low frequency regime, and by IXS in the mesoscopic regime, is reported in Fig. 2, along with the photoinduced differential reflectivity oscillations map and IXS spectra at fixed exchanged momenta (insets a and b), from which the dispersion is obtained.
In the low-frequency region, the linear energy dispersion
indicates the existence of propagating vibrational excitations
corresponding to a single longitudinal acoustic-like phonon branch.
Remarkably, the propagating nature of these excitations
lingers up to the THz frequency regime. The presence of structural disorder, indeed, strongly attenuates acoustic modes at the nanometer scale. Nevertheless, the existence of a well defined first sharp diffraction peak mimics a first pseudo-Brillouin zone and, though the vibrational eigenmodes are no longer plane waves, a well defined dominant wavevector still exists up to a zero-group velocity point corresponding to half of the pseudo-Brillouin zone boundary \cite{sette_science}.
The obtained sound velocity (the $q=0$ derivative of the best sinusoidal fit to IXS data, dashed line in Fig. 2) is in excellent agreement with BPA data in the ten GHz range and with previous single (lower) frequency determination \cite{kearns_high-modulus_2010}.

The existence of a well defined dispersion at any lengthscales within a pseudo-Brillouin zone, suggests the potential sensitivity of the acoustic properties to the local topology of the inherent structures visited by the glass.
Given its capability to access a sizable portion of the dispersion curve, determining sound velocity with superior accuracy (down to 0.1 $\%$, about one order of magnitude better than ordinary Brillouin Light Scattering) BPA measurements have been extended to the other PVD samples in Fig. 3 (a and b). During the experiments the temperature was kept constant at T=295 K, i.e. well below $T_g$ and above the $T_f$ of the different depositions, to ensure that the glass state is arrested on the laboratory timescale.

A clear inverse correlation between sound velocity and $T_{f}$ arises. BPA experiments also enable to determine the hypersonic attenuation $\Gamma$ with an accuracy of tenths of GHz, and, accordingly, determine the frequency dependence of the life-time of the vibrational excitations. As an example, the frequency dependence of $\Gamma$ for $T_f$ = 281 K and 301 K are reported (pink and blue, respectively) in Fig. 3 (c). Aiming to unravel any link with $T_f$, we integrated the attenuation over the explored frequency range and calculated the excess relative (and normalized) to the most ultrastable glass ($T_{f}=281$ K), as reported in Fig. 3 (d). The relative attenuation shows, similarly to the sound velocity, a clear but opposite (direct) correlation with $T_f$. Acoustic attenuation in disordered materials is ruled by different physical mechanisms \cite{natcom}. In the  GHz regime, where a crystal-like picture of vibrational excitations holds, the attenuation is expected to be due to the anharmonicity of the interparticle interaction. The decrease of sound attenuation towards the lowest $T_{f}$'s discussed above, therefore, corroborates the idea that lower energy basins are more harmonic in nature.
\section{Discussion}
The distinct thermo-mechanical correlation reported in this study has important implications for the validity of those paradigms which can be uniquely benchmarked at the liquid side of the glass transition, since equilibration times dramatically increase below $T_g$ preventing direct explorations.
Fragility, in particular, quantifies the steepness of the liquid viscosity (or equivalently the relaxation time, through Maxwell's relation $\eta=G_\infty \tau$) at $T_g$, and it has a pivotal role in controlling physical properties of the supercooled phase \cite{BANGALORE,angell_sup}. Remarkably, the provocative idea that fast degrees of freedom characterizing glass dynamics well below $T_g$ can be predictors for slow dynamical properties of the liquid state, such as the structural relaxation time and hence fragility, is rapidly gaining consensus \cite{kai,buchenau,scopigno_fragility,frag,larini}. While density fluctuations completely decorrelate in a liquid, as a consequence of the ergodic sampling of different basins in the energy landscape, in a glass, a residual correlation exists, uniquely determined by the entire spectrum of vibrational eigenstates of the energy minimum where the system is trapped. This is the so called non ergodicity factor (NEF), the long time limit of the density-density autocorrelation function.
Of interest here, it has been shown that the liquid fragility, m, can be determined by the low temperature behavior of the NEF, $f_q(T)$ \cite{scopigno_fragility,frag}. The NEF can be quantified in different ways. Among them, by the relative sound velocity jump occurring at the glass transition \cite{niss,buchenau_contro}, namely:
\begin{equation}
m=\gamma\cdot(f^{-1}(T_g)-1) \approx 140\frac{c^2_0}{c^2_{\infty}-c^2_0}
\label{alfa}
\end{equation}
where $c_0$ and $c_{\infty}$ are the liquid and glass sound velocity values, respectively, and $\gamma \approx 140$ expresses the above discussed correlation with fragility m \cite{scopigno_fragility}, verified in ordinary IMC glass \cite{pogna}.
Critically, $c_0$ is an equilibrium property of the liquid, while $c_{\infty}$ is shown here to depend upon the very stability of the corresponding glass. This advocates the extension of the fragility concept to a temperature dependent steepness index \cite{jeppe_natphys}, $I(T)$ such that $I(T_g) = m$, which we use here to obtain the equilibrium viscosity below $T_g$. The mechanical response of a glass of a given stability, indeed, determines the viscosity of its liquid at the corresponding $T_f$, via the dependence $I(T=T_f)=I(c_{\infty}(T_f))$ in Eq. \ref{alfa}, reported in Fig. 4 (a).
Integrating this latter (see Eq. \ref{notes} in Materials and Methods), an Arrhenius plot can ultimately be obtained, shown in Fig. 4 (b). The super-Arrhenius behaviour documented in IMC above $T_g$ \cite{descamps} is reported as dashed line representing a Vogel-Fulcher-Tamman (VFT) function. For $T < T_g$ a remarkable deviation is observed: the apparent fragility is larger than m, but falls below the VFT expectation, and decreases with the glass stability, signifying a fragile-to-strong transition, quantified in Fig. 4 (a). This result verifies the predictions of Kovacs and Adam-Gibbs-Vogel models from aged IMC \cite{brian} and rationalizes very recent work in naturally long-time aged amber \cite{mcK}, which set an upper bound to the temperature dependence of the dynamics below $T_{g}$. It is also in line with the conclusion against the VFT extrapolations in the glass \cite{jeppe_natphys,mauro_pnas}, and relates to the observation of an additional non Arrhenius equilibration process in polymers \cite{cangialosi}.
Moreover, the simultaneous decrease of both the acoustic attenuation and generalized fragility reported here at low $T_{f}$'s, validates recent molecular dynamic simulations \cite{ngai} which put forward a direct correlation between kinetic fragility m and the degree of anharmonicity of the interparticle interaction potential.

All together, these evidences syncretize on the scenario schematically depicted in Fig. 4, establishing a firm link between the hypsometric characterization of the energy landscape and basin-specific vibrational properties. Crucial to technological applications, the mechanical approach advanced here to assess stability is a non destructive one, as opposed to calometric determinations of $T_f$. Most important, we demonstrate that the state of the glass is totally identified by $T_f$ and $T$, and once the non ergodicity parameter is calculated by the sound velocity jump at $T_f$ it is possible to determine the relaxation time from equation Eqs. (\ref{alfa}) and (\ref{notes}), as shown in Fig. (\ref{f:fragility}). The emerging protcol provides a unique way to capture essential features of a liquid attaining the structural arrest away from $T_{g}$, up to relaxation times beyond $10^{9}$ s, corresponding to exapoise viscosities.
In conclusion, we circumvented here the major hindrances to probe equilibrium properties below the ordinary glass transition temperature: crystallization tendency and unrealistic observation times. The isothermal measurement of mechanical properties for glassy phases with different $T_f$ reveals the increase of the longitudinal elastic modulus and decrease of the acoustic loss below $T_g$, which we connect to a breakdown of the VFT law. This may lead to reconsider old paradigms based on dynamical divergency to ever reach a definitive understanding of the glass transition.

\section{materials and methods}
\subsection{Sample growth and thermal characterization and treatment}
Indomethacin (IMC, 99$\%$ purity, $T_{g}$= $315$ K and $T_{m}$ ($\gamma$ form)= $428$ K) crystalline powders were purchased from Sigma-Aldrich.
Twin films of ultrastable IMC glasses were grown by vapor deposition on Si (100) substrates and on differential scanning calorimetry Al pans at deposition temperatures spanning from 190-to-290 K. An effusion cell filled with IMC was heated to achieve the desired deposition rate of 0.1 nm/s, as measured by a quartz crystal microbalance. When this rate was attained, the shutter was removed to start deposition. The thickness of the films ranges from 1.5-2 $\mu$m for the BPA experiments to 50 $\mu$m for the IXS measurements. All samples were stored in vacuum-sealed bags with desiccant in the freezer to minimize aging prior to the acoustic measurements, which were carried out several days after the deposition. Differential scanning calorimetry was performed immediately after removing the samples from the growth chamber \cite{insitu}. A DSC Perkin Elmer 7 was used to monitor the power absorbed/released during heating scans at a rate of 10 K/min on IMC films of 2-10 mg. The first scan typically corresponds to an ultrastable glass, while the second one is characteristic of a glass obtained by cooling the liquid at -10 K/min. The fictive temperature is measured, evaluating the intersection of the enthalpy curves of the PVD glasses, i.e. the integral of the heat capacity upscans, with the supercooled liquid line extrapolated according to a quadratic fit \cite{kearns_hiking}. Notably, the extrapolated value of the fictive temperature does not depend on the heating rate and directly reflects the structural state of the glass. The lower values of the fictive temperature of ultrastable glasses signify a smaller enthalpy content compared to the glasses cooled at ordinary rates.

\subsection{Inelastic x-ray scattering}
The IXS experiments were performed at the beam line ID28 of the European Synchrotron Radiation Facility (ESRF).
The experimental observable is the frequency spectrum of the scattered X-ray intensity, proportional to the dynamic structure factor $S(q,\hbar \omega$), where $q$ is the exchanged momentum defined by the scattering angle $\theta$ and by the wavevector of the incident photons $k_i$ as $q=2k_i sin (\theta /2)$. A 8-analyzers bench was utilised, operating in horizontal scattering geometry. The sample was mounted with the substrate parallel to the scattering plane, tightly focusing the beam down to ~$20 \mu$m in the vertical direction to avoid parasitic scattering from the substrate. With such a geometry the probed phonon direction is orthogonal to the growth.
Energy scans were performed at constant $q$ values in the range $1$ to 8 $\approx$ nm$^{-1}$, corresponding to half of the pseudo Brillouin Zone (q=$14.7$ nm$^{-1}$). The $q$ resolution was determined by slits placed in front of the analyzer was set to $0.25$ nm$^{-1}$.
The scanned energy range was $-30~\leq \hbar \omega\leq~30$ meV, where $\hbar \omega=E_0 - E$ is the energy transfer, with $E_0$ and $E$ being the energy of the incident ($23.725$ eV) and the scattered x-ray photon; each scan took approximately $480$ min. Using the (12 12 12) reflection for the Si monochromator and crystal analyzers the overall energy resolution was $1.5$ meV the FWHM of instrumental function (black line under the spectra in Fig. 2). Measurements were performed at room temperature (T= 295 K) on a sample grown by PVD at substrate temperature T=266 K and deposition rate of 0.1 nm/s, $50\mu$m thick.
The dynamic structure factor, $S(q,\hbar \omega$), contains information about the sound dispersion $\Omega (q)$ and attenuation $\Gamma (q)$ which can be extracted by a damped harmonic oscillator model (DHO):
\begin{equation}
\begin{split}
S(q,\omega)=S(q)\left [\delta(\omega)f_{q}+ \frac{(1-f_q)}{\pi}\frac{\Omega(q)^2\Gamma(q)}{[\omega^2-\Omega(q)^2]^2+\omega^2\Gamma^2(q)} \right ] \\
\cdot\frac{\hbar\omega/(k_{B}T)}{1-e^{\hbar\omega/(k_{B}T)}}
\end{split}
\label{fitS}
\end{equation}
were the Bose factor accounts for the quantum nature of the probed excitations. The two terms in Eq. \ref{fitS} represent the elastic and the inelastic components of the spectra, respectively. Eq. \ref{fitS} was fitted to the experimental data after a correction for the finite instrument resolution \cite{scopigno_microscopic_2005}.

The IXS sound velocity measurement of the ordinary glass reported in Fig. 3 (b) accounts for a 0.8$\%$ positive dispersion occurring in the THz regime \cite{pogna}.

\subsection{Picosecond photoacoustics}
Time-domain measurements of longitudinal sound velocity and acoustic damping in the hydrodynamics limit
were performed by broadband picosecond photoacoustics.
This pump and probe technique is based on the generation and detection of coherent vibrational excitations
by means of ultrashort laser pulses. The setup is built on a regeneratively amplified Ti:sapphire laser producing
50 fs, 4 mJ pulses at 800 nm with 1 kHz repetition rate. The output is split to generate both the pump and the probe beams; the former, after passing through a delay
line, is focused onto the sample on a $\approx$ 100 $\mu$m spot diameter with energies up to 5 $\mu$J. The PVD glasses of IMC, 1.5 $\mu$m $\div$ 2 $\mu$m thick, have been deposited onto a crystalline Si substrate at six different substrate temperatures (T= 190, 210, 236, 266, 285, 290) and coated by a 15 nm thick layer of Ni. As the pump impinges on the sample, it is partially absorbed by the metallic coating, which thermally expands launching very short strain pulses, i.e. longitudinal acoustic wave packets in the underlying glass, with a characteristic spectrum extending from few to hundreds of GHz \cite{thomsen}.

The photo-generated acoustic wave packet travels inside the sample along the growth direction. A portion of the probe pulse is reflected at the metallic surface, while the transmitted component interferes with the light scattered from the traveling density fluctuation associated with the acoustic wave packet. By monitoring transient differential reflectivity (with and without the pump pulse) $\frac{\Delta R (t)}{R}$ as function of the time pump-probe time delay, information can be gained on the phonon spectrum. Specifically, at any given probe wavelength $\lambda$, the phonon frequency is set by the Bragg condition corresponding to a stimulated Brillouin scattering event:
\begin{equation}
\nu=\frac{v}{2\pi}q=\frac{2v_{s}}{\lambda}\sqrt{n^2-\sin^2{\beta}}
\label{phase_match}
\end{equation}
where $v$ is the longitudinal sound velocity, $q$ the exchanged momentum, $n(\lambda)$ the index of refraction of the glass, and $\beta$ is the incident angle with respect to the metallic coating.
In our realization, the probe pulse consisted of a broadband white light continuum, generated from 2 mm thick Sapphire crystal plate and filtered to select wavelengths in the range 450 $\div$ 650 nm. The broadband probe pulse's reflection was dispersed by a 150 g/mm grating and monitored by a CCD, such that a number of two color pump and probe experiments were simultaneously recorded. Transient reflectivity data were first reduced by subtraction of an exponential thermal background. Each channel achieves a sensitivity up to $\frac{\Delta R}{R} \approx 10^{-5}$. In order to get rid of an independent determination of the refractive index n, required in Eq. \ref{phase_match}, the measurements were repeated at different scattering angles $\beta$. A selection of the reflectivity oscillations as function of time and probe wavelength is reported in color map in Fig. 2 together with the signals at selected wavelengths. The signal was modeled with the damped harmonic oscillator in time domain, whose frequency can be estimated with an accuracy of 0.1 $\%$. The sound velocity was determined from the linear fit of the modes' frequency $\nu$ as function of the exchanged momentum $q$ over all the broadband probed range. The acoustic attenuation, indeed, was directly extracted from the damping of the reflectivity oscillations with 6 $\%$ accuracy.

\subsection{Temperature dependence of the relaxation time}
The so called kinetic fragility quantifies the temperature behaviour of the relaxation time $\tau (T)$ at a conventional fictive temperature ($T_g$) where $\tau$ =100 s. Namely:
\begin{equation}
m=\left.\frac{\partial log_{10} \tau}{\partial (T_g/T)}\right |_{T=T_g}
\end{equation}
In order to describe the low temperature dependence of the non ergodicity factor $f_q(T)$ in the low $q$ limit, a dimensionless steepness index can be introduced:
\begin{equation}
\alpha=-\left.\lim_{q\rightarrow 0}\frac{\partial f}{\partial\frac{T}{T_g}}\right |_{T=0}
\end{equation}
Remarkably, $\alpha$ exhibits a strong direct correlation with the kinetic fragility m, as it has been demonstrated experimentally \cite{scopigno_fragility,frag,greer,mauro_prb} and by numerical simulations \cite{ngai} for a sizable number of glass-formers. Accordingly, the faster is the T dependence of $f_q(T)$, i.e. the higher is the value of $\alpha$, the more fragile is the glass-former. Below fragilities as high as one hundred, the correlation is linear:
\begin{equation}
m=\gamma \alpha
\end{equation}
with $\gamma \approx 140$.
In the harmonic approximation, the T-dependence of $f_q(T)$ reads:
\begin{equation}
f_q(T)=\frac{1}{1+\alpha\frac{T}{T_{g}}}
\label{nef}
\end{equation}
The value of $f_q(T_g)$ can be conveniently determined from the glass ($c_{\infty}$) $\rightarrow$ liquid ($c_0$) sound velocity jump around the glass transition, through the expression:
\begin{equation}
f_q(T_g)=\left. 1-\frac{c_{o}^2}{c_{\infty}^2}\right |_{T_g}
\label{nef1}
\end{equation}
By evaluating Eq. \ref{nef} at $T=T_g$, and combining it with Eq. \ref{nef1}, it immediately follows:
\begin{equation}
m=\gamma \left. \frac{c^2_0}{c^2_{\infty}-c^2_0}\right |_{T_g}
\label{alfa_sm}
\end{equation}
Beside the measurement temperature, however, the glass velocity also depends upon stability, i.e. on the fictive temperature $T_f$. This can be taken into account by generalising Eq. \ref{alfa_sm} as:
\begin{equation}
\left.\frac{\partial log_{10} \tau}{\partial (T_f/T)}\right |_{T=T_f} = \gamma \left.\frac{c^2_0(T)}{c^2_{\infty}(T_f,T)-c^2_0(T)}\right |_{T=T_f} = I(T_f)
\label{gen_frag}
\end{equation}
where the notation $c_{\infty}(T_f,T)$ explicitly indicates the above mentioned dependencies on fictive and measurement temperatures, respectively. It is worth mentioning that the T-dependence, in Eq. \ref{gen_frag} (i.e. on the very temperature where the sound velocity jump is evaluated) is negligible compared to the $T_f$ dependence brought about by the glass sound velocity.
Once integrated in $1/T$, Eq. \ref{gen_frag} provides the relaxation time below $T_g$ through:
\begin{equation}
log_{10} \tau (1/T_f)=log_{10} \tau (1/T_g) + T_f \int_{T_g^{-1}}^{T_f^{-1}} I(T') d(1/T')
\label{notes}
\end{equation}
with $c_0(T_g)$ = 1354 m/s estimated by Brillouin Light Scattering \cite{pogna_pccp}, and the $c(T)$ dependence is from \cite{kearns_high-modulus_2010}.
In Fig. 5 we sketch the evolution of the sound velocity as function of temperature across the glass transition, for two glasses of different fictive temperature.

\section{Author contributions}
T.S. conceived and supervised the research. C.R.-T. and J.R.-V. prepared the samples and carried on the calorimetric characterization. E.A.A.P. and C.F. performed the Picosecond Photo-Acoustics measurements and data analysis. E.A.A.P, C.R-T and T.S. performed the IXS experiment. E.A.A.P. and T.S. wrote the manuscript. All the authors participated to the discussion of the results and commented on the manuscript.

\begin{acknowledgments}
E.A.A.P., C.F. and T.S. have received funding from the European Research Council under the European Community's Seventh Framework Program (FP7/2007-2013)/ERC grant Agreement No. 207916. C.R.T and J.R.V. acknowledge financial support from Generalitat de Catalunya and Ministerio de Economía y Competitividad through grants SGR2009-01225 and MAT2013-40896-P, respectively.  We thank M. I. Alonso and M. Garriga from ICMAB for the ellipsometric measurements, and the ID28  staff for the support during the IXS experiment.
\end{acknowledgments}

\bibliographystyle{naturemag}


\begin{figure}
\centering
\includegraphics[width=0.95\textwidth]{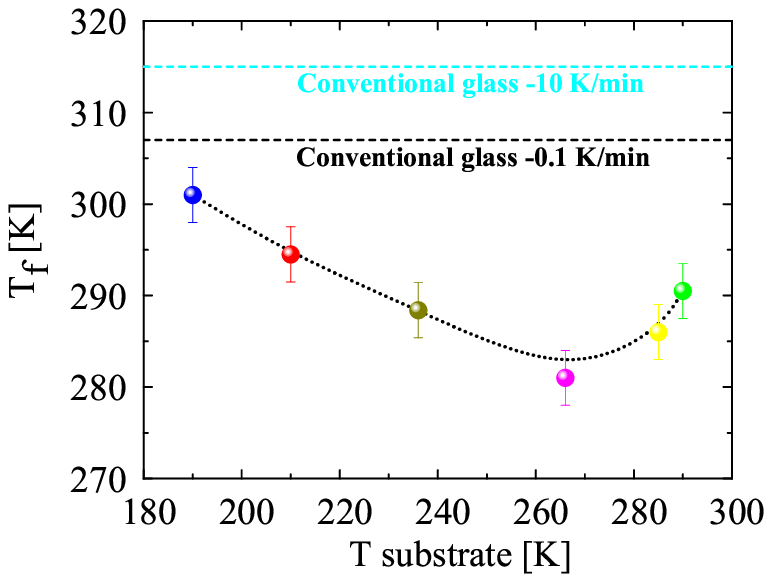}
\caption{
Fictive temperature of PVD glasses as function of the substrate temperature during deposition, determined from DSC upscans, directly relates to the enthalpy content, hence quantifying the glass stability (see Materials and Methods for more details). Typical values obtained for conventional glasses for two different cooling rates are also indicated (dashed lines)}
\label{f:tonsets}
\end{figure}

\begin{figure}
\centering
\includegraphics[width=0.95\textwidth]{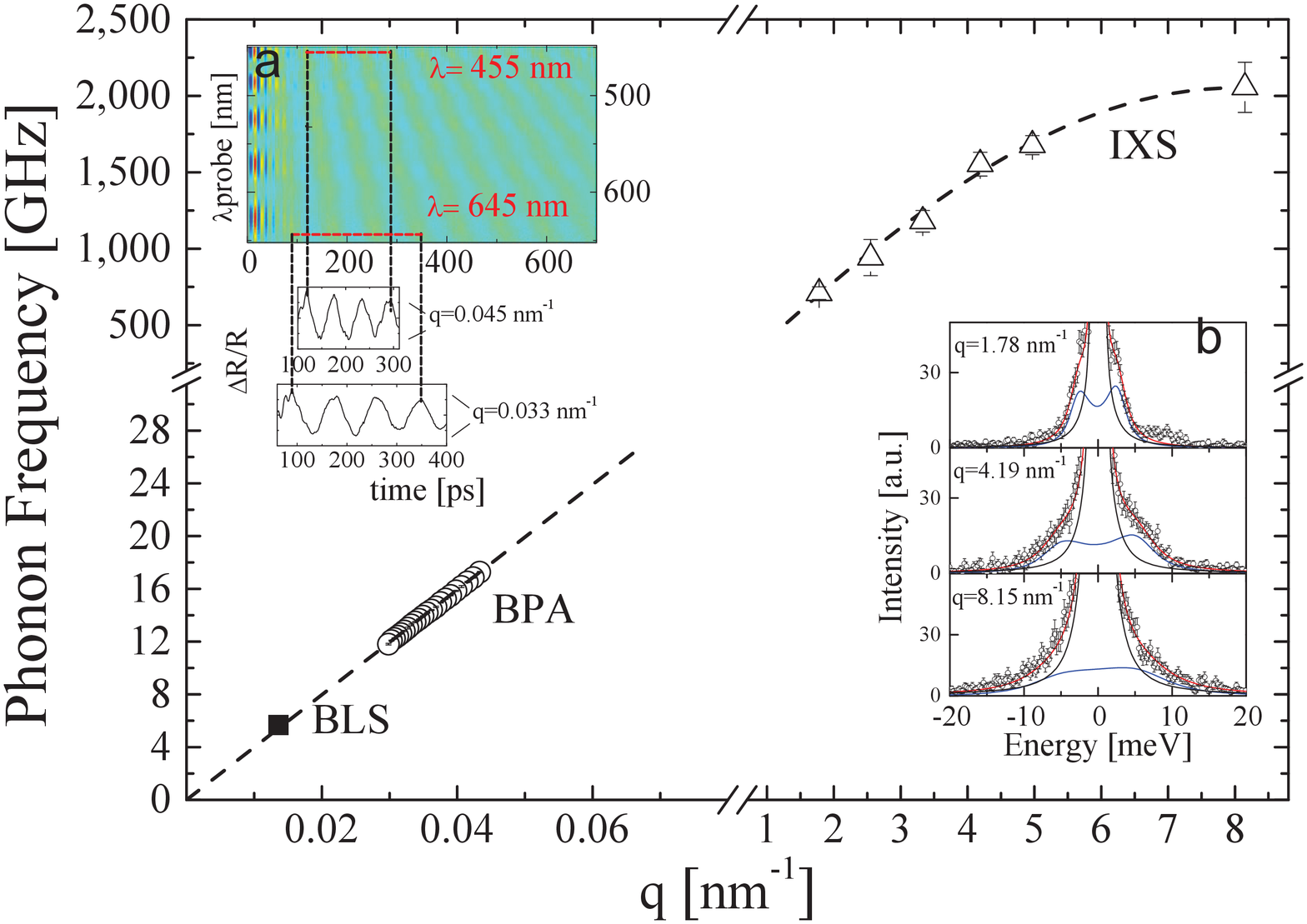}
\caption{
Longitudinal acoustic phonon dispersion in ultrastable PVD glass ($T_{sub}$=266 K), investigated by IXS ($\triangle$, dashed line is the sinusoidal best fit) in the THz, by BPA ($\circ$) in the GHz regime, along with a previous Brillouin measurement ($\Box$) at lower frequency \cite{kearns_high-modulus_2010}. (\textbf{a}) Transient reflectivity BPA data, each $\lambda_{probe}$ couples to a phonon of well defined wavevector, visualized as an oscillatory signal in time domain. (\textbf{b}) IXS spectra, acoustic phonons for selected wavevectors, $Q$, probed in frequency domain.}
\label{f:dispersion}
\end{figure}
\begin{figure}
\centering
\includegraphics[width=0.95\textwidth]{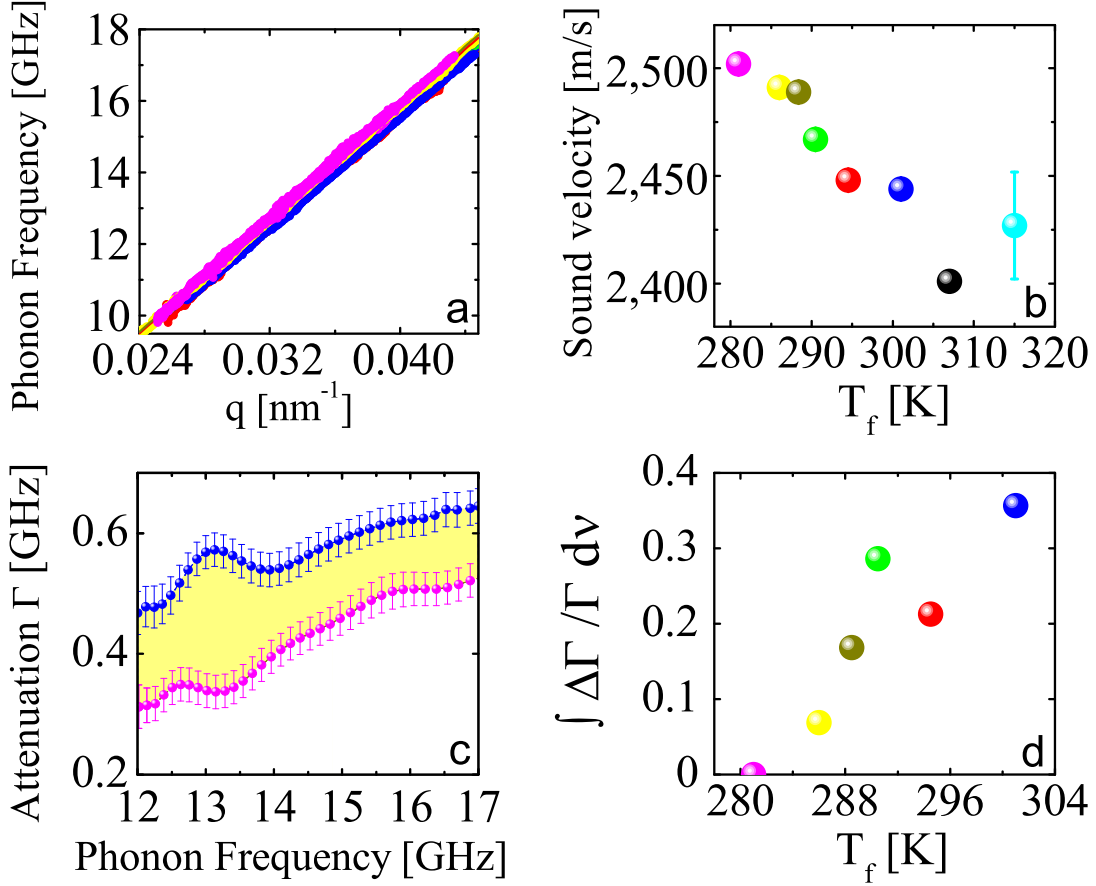}
\caption{
Sound velocity and damping for different stabilities.
(\textbf{a}) GHz-dispersion curves of the longitudinal acoustic phonons from isothermal BPA measurements ($T=295$ K).
(\textbf{b}) Sound velocity (linear fit to the dispersion curves), exhibits a correlation with $T_{f}$, showing a maximum for $T_{f}=281$ K. Error bars smaller than the symbol. The values for conventional glasses ($T_{f}=307$ K and $T_{f}=315$ K) are from \cite{kearns_high-modulus_2010} and \cite{pogna}, respectively.
(\textbf{c}) Acoustic attenuation, $\Gamma$, for depositions with high ($T_{sub}=266$ K, pink) and low ($T_{sub}=190$ K, blue) stabilities as function of the phonon frequency. (\textbf{d}) Normalized excess acoustic attenuation relative to the most ultrastable glass, obtained integrating over the explored frequency range and normalizing to the $T_{f}=281$ K case. An inverse correlation respect to that in sound velocity is observed.}
\label{f:velocity}
\end{figure}

\begin{figure}
\centering
\includegraphics[width=0.95\textwidth]{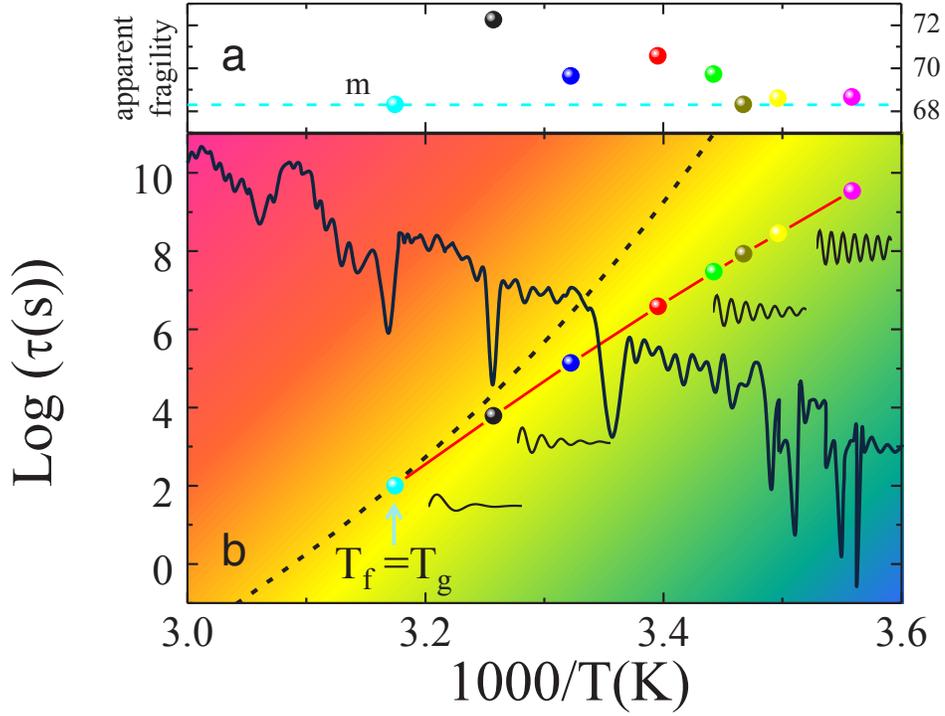}
\caption{
Viscous flow of a liquid and stability of its glasses.
(\textbf{a}) Apparent fragility $I(T)$, determined from the sound velocity jump in glasses with different $T_f$, indicates a fragile to strong transition below $T_g$. Conventional fragility $m=I(T_g)$ is also reported. (\textbf{b}) Arrhenius plot for $T<T_g$, obtained integrating data from panel (a). A correlation is found between the depth of the inherent structures in the energy landscape (grey line) probed by glasses of different $T_f$ and the mechanical response (sketched by time domain damped oscillations). The VFT behaviour obtained from above $T_g$ data \cite{descamps} is also shown (dashed line).}
\label{f:fragility}
\end{figure}
\begin{figure}[ht]
\centering
\includegraphics[width=0.95\textwidth]{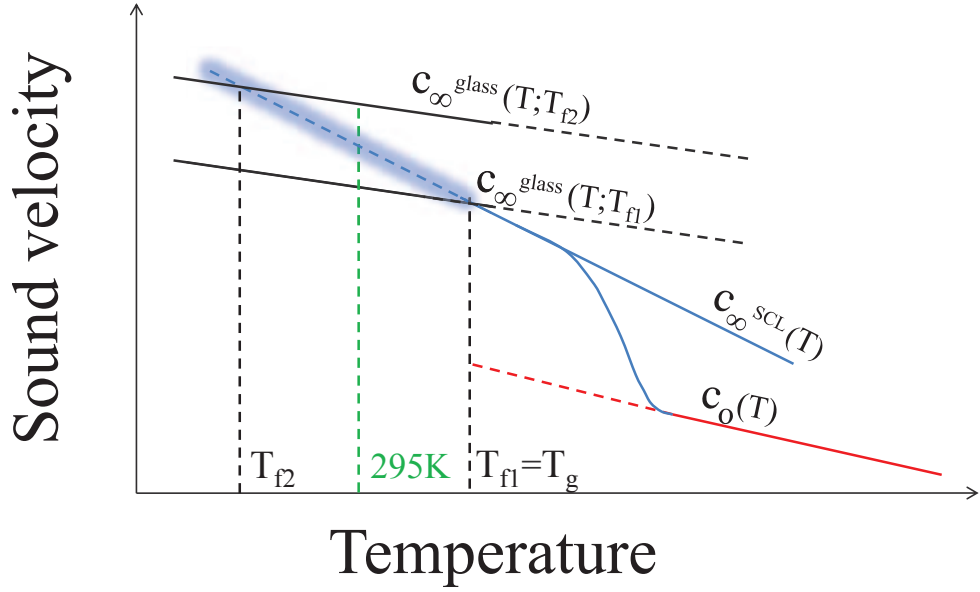}
\vspace{-2cm}
\caption{
Temperature dependence of the sound velocity for two glasses of different fictive temperature, $T_f$, along with the unique liquid and supercooled liquid values. The shaded line indicates the region of the highly viscous supercooled liquid, where direct observations are prevented by impractically long equilibration timescales.}
\label{f:cinf}
\end{figure}

\end{document}